\documentclass[prb,a4paper,twocolumn,showpacs]{revtex4-1}
	
\usepackage[utf8]{inputenc}
\usepackage[T1]{fontenc}
\usepackage{lmodern}

\usepackage{amssymb}
\usepackage{amsmath}
\usepackage{amsbsy}
\usepackage{bm}
\usepackage{graphicx}

    \newcommand{\I}{\mathrm{i}}
    \newcommand{\E}{\mathrm{e}}

\def\Xint#1{\mathchoice
{\XXint\displaystyle\textstyle{#1}}%
{\XXint\textstyle\scriptstyle{#1}}%
{\XXint\scriptstyle\scriptscriptstyle{#1}}%
{\XXint\scriptscriptstyle\scriptscriptstyle{#1}}%
\!\int}
\def\XXint#1#2#3{{\setbox0=\hbox{$#1{#2#3}{\int}$ }
\vcenter{\hbox{$#2#3$ }}\kern-.58\wd0}}
\def\fint{\Xint-}

\begin{document}

\title{Dynamical spin Hall conductivity in a magnetic disordered system}
\author{T. L. van den Berg}
\email{tineke.vandenberg@im2np.fr}
\affiliation{Aix-Marseille Université, IM2NP-CNRS UMR 6242, Campus St. Jérôme, Case 142, 13397 Marseille, France}
\author{L. Raymond}
\email{Laurent.Raymond@univ-provence.fr}
\affiliation{Aix-Marseille Université, IM2NP-CNRS UMR 6242, Campus St. Jérôme, Case 142, 13397 Marseille, France}
\author{A. Verga}
\email{Alberto.Verga@univ-provence.fr}
\affiliation{Aix-Marseille Université, IM2NP-CNRS UMR 6242, Campus St. Jérôme, Case 142, 13397 Marseille, France}

\date{\today}

\begin{abstract}
We investigate the intrinsic spin Hall effect in a quantum well semiconductor doped with magnetic impurities, as a means to manipulate the carriers' spin. Using a simple Hamiltonian with Rashba spin-orbit coupling and exchange interactions, we analytically compute the spin Hall conductivity. It is demonstrated that using the appropriate order of limits, one recovers the intrinsic universal value. Numerical computations on a tight-binding model, in the weak disorder regime, confirm that the spin Hall effect is preserved in the presence of magnetic impurities. The optical spin conductivity shows large sample to sample fluctuations in the low frequency region. As a consequence, for weak disorder, the static spin conductivity is found to follow a wide Gaussian distribution with its mean value near the intrinsic clean value.
\end{abstract}


\maketitle

\section{Introduction}
\label{sec.intro}
The possibility to manipulate the electron spin by electric currents was first proposed by Dyakonov and Perel, forty years ago.\cite{Dyakonov-1971fr} They noted that, in analogy with the anomalous Hall effect, an electric current must create a spin flow perpendicular to the charge current. This ``spin Hall'' effect is a consequence of the asymmetry, driven by the disorder potential through spin-orbit interaction, in the spin and momentum of the electrons scattered off impurities. Renewed interest in this effect arose more recently, when an intrinsic counterpart was predicted in strongly spin-orbit coupled semiconductors.\cite{Murakami-2003pb,Sinova-2004fb} The intrinsic effect results from the spin-orbit coupling in the underlying topology of the Bloch states (in momentum space), that gives rise to an anomalous component in the group velocity, perpendicular to the applied electric field.\cite{Murakami-2006df,Inoue-2009fk} It is based entirely on the one-particle band structure, and therefore subject to the influence of disorder. In the case of an n-doped semiconductor quantum well, the intrinsic contribution to the spin Hall conductivity turns out to be exactly cancelled by the effect of scattering.\cite{Schwab-2002vn,Mishchenko-2004ys,Inoue-2004fk,Dimitrova-2005uq,Chalaev-2005fk} The exact cancellation of the intrinsic spin Hall effect is particular to the Rashba spin-orbit coupling, linear in momentum, and is not present for instance in the Luttinger model of a p-type semiconductor.\cite{Murakami-2004kx} 

Experimental observation of the spin Hall effect was achieved first by optical techniques to detect the spin accumulation,\cite{Kato-2004fu,Wunderlich-2005dq} and more recently by direct all-electrical measurements.\cite{Garlid-2010fj} The inverse intrinsic effect, that is the generation of a longitudinal charge current by the injection of polarized electrons has also been reported.\cite{Werake-2011fk}

\begin{figure}[!b]
\centering%
\includegraphics[width=0.48\textwidth]{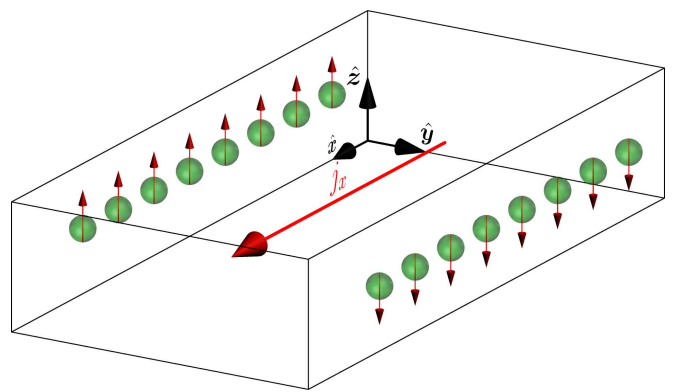}
\caption{(Color online) Schematic geometry of the spin Hall effect; a longitudinal current \(j_x\) driven by an external electric field induces a transversal spin current that results in a polarization of the carrier spins following the sign of their momentum (spheres with arrows).}
\label{fig.hall-V}
\end{figure}

The vanishing of the spin conductivity in the Rasbha model of a two-dimensional electron gas, for arbitrarily small disorder, can be traced back to the fact that the spin current operator \(\bm J^z=(\hbar/4)\{\sigma^z,\bm v\}\), is proportional to the spin \(\bm s\) derivative 
\[
J^z_y\sim\dot{s}_y=\frac{\I}{\hbar}[H,s_y]\,,
\]
where \(\bm\sigma=(\sigma_x,\sigma_y,\sigma_z)\) is the vector of the Pauli matrices, \(\bm v\) is the electron velocity, \(H\) is the Hamiltonian, and \((x,y,z)\) refer to the directions of the applied electric field, the generated transverse spin current, and the out of plane spin polarization (Fig.~\ref{fig.hall-V}). The commutator \([H,s_y]\) is not affected by the presence of randomly distributed impurities. In a stationary state, the time derivative of the averaged spin density vanishes, and hence the spin current must also vanish.\cite{Dimitrova-2005uq,Chalaev-2005fk,Rashba-2004zr} However, if the impurities were magnetic this argument is no longer valid, the spin current would depend on additional contributions, and could therefore remain finite. As a consequence, magnetic impurities should in principle be able to restore the intrinsic Hall effect in a two-dimensional n-doped semiconductor. 

The influence of magnetic impurities on the intrinsic Hall effect was investigated using numerical computations of the Landauer formula,\cite{Liu-2006fk} linear response theory and Kubo formula,\cite{Inoue-2006fk,Wang-2007uq,Moca-2007tg} and quasiclassical Green functions.\cite{Gorini-2008kx} It is well established that, contrary to the non-magnetic case, the spin Hall effect does not vanish in this case, although different values of the spin conductivity in the clean limit were found.\cite{Inoue-2009fk} Interest in materials combining the electronic properties of semiconductors with ferromagnetism was enhanced by their potential applications in spintronics.\cite{Fabian-2007kx,dietl2008spintronics,Sato-2010fk} The main focus in the study of the so-called ``diluted magnetic semiconductors'' is on the mechanisms governing the interactions between impurities, that determine their thermodynamical and magnetic properties.\cite{Jungwirth-2006ca} However, it would also be interesting to investigate the spin dependence of the carriers response, to an applied electric field. Indeed, while the topological component of the anomalous Hall effect is widely documented in magnetic semiconductors, its spin current counterpart, the intrinsic spin Hall effect, is still a developing topic.\cite{Nagaosa-2010vn} This is the question we address in the present paper. Our main goal is to compute, using linear response theory, the intrinsic spin conductivity in a confined electron gas taking into account Rashba spin-orbit coupling and magnetic disorder. In particular, we want to obtain an analytical expression of the dynamical spin conductivity, to calculate first the clean limit for fixed frequency, and then the static limit. The order of limits is important here because the large scattering time limit does not commute with the small frequency limit.\cite{Murakami-2004kx}

In Sec.~\ref{sec.system}, we consider a simple model suitable to describe the linear response of a confined electron gas to an external electric field, as determined by the spin-orbit coupling (inversion asymmetry) and scattering on magnetic impurities (assumed to be in a paramagnetic state). We neglect extrinsic effects and non-magnetic impurities, and take into account an effective mass conduction band. In Sec.~\ref{sec.numerics} we generalize the model to a finite band tight-binding Hamiltonian in a square lattice. As in the effective mass model, we consider spatially distributed magnetic impurities with random orientations. We use a Chebyshev pseudo-spectral method to compute the density of states (total and local) and the spin Hall conductivity.\cite{Weisse-2006fk} Finally, we discuss and summarize our results (Sec.~\ref{sec.conclusion}).

\section{Spin Hall dynamical conductivity}
\label{sec.system}
In this section we compute the spin conductivity \(\sigma_{sH}(\omega)\) as a function of the frequency \(\omega\), starting with a simple model Hamiltonian \(H\). The Kubo formula is analytically calculated in the weak disorder limit, taking into account the ladder diagrams.

\subsection{Model Hamiltonian}
\label{subsec.model}
We consider a two-dimensional system of noninteracting electrons of effective mass \(m\), charge \(-e\), and Rashba spin-orbit coupling constant \(\lambda\), in the presence of randomly distributed magnetic impurities,
\begin{equation}
\label{Hcont}
H = \frac{{\bm p}^2}{2m} - 
    \frac{\lambda}{\hbar}{\bm \sigma}\cdot(\hat{z}\times {\bm p}) +
    V(\bm x)\,.
\end{equation}
where \(\bm p\) and \(\bm x\) are the momentum and position operators of the electron in the plane \((x,y)\). The impurity potential energy is assumed to be short ranged and characterized by an exchange interaction constant \(J_s>0\) and a microscopic length \(a\),
\begin{equation}
\label{Vm}
V(\bm x)=J_s a^2\sum_{i\in I} \hat{\bm n}_i\cdot \bm\sigma\, 
     \delta(\bm x - \bm x_i)
\end{equation}
where \(I\) is the set of impurity sites \(\bm x_i\), and \(\hat{\bm n}_i\) the unit vector in the direction of the impurity magnetic moment; we assume that the impurities are in a paramagnetic state, so \(\hat{\bm n}_i=(\sin \theta_i \cos\phi_i,\sin \theta_i \sin\phi_i, \cos\theta_i) \) is uniformly distributed over the sphere, with \(\theta_i\in(0,\pi)\) and \( \phi_i \in (0,2\pi) \) independent random variables. Typical orders of magnitude, for a III-V semiconductor heterostructure, are: effective mass \(m = 0.067\, m_e\) (\(m_e\) is the electron mass), lattice length \(a\sim 0.56\,\mathrm{nm}\), Rashba constant \(\lambda\sim 10^{-11}\,\mathrm{eV\,m}\), and s-d exchange constant (for Mn impurities, for example) \(J_s\sim 0.1\,\mathrm{eV}\).\cite{Fabian-2007kx,Wu-2010fk} It is convenient to choose units such that \(\hbar=m=a=e=1\).

The clean Hamiltonian becomes diagonal in the chiral base,\cite{Arii-2007kx}
\begin{equation}
\label{eqn.eigenS}
|\alpha_\pm \rangle = \frac{1}{\sqrt{2}}
    \begin{pmatrix} 
        \pm \I \E^{- \I \varphi}\\ 
        1
    \end{pmatrix}\,,
\end{equation}
where \( (k,\varphi) \) are polar coordinates in momentum space \(k_x+\I k_y=k\E^{\I \varphi}\). The corresponding eigenvalues are,
\begin{equation}
\label{eqn.eigenE}
\epsilon_{\pm}(k) = \frac{k^2}{2}\pm \lambda k\,.
\end{equation}
The transformation matrix between the spin and the chiral basis is 
\begin{equation}
\label{eqn.u}
U (\varphi) = \frac{1}{\sqrt{2}} 
    \begin{pmatrix} 
        - \I \E^{- \I \varphi} & \I \E^{- \I \varphi} \\
        1 & 1
    \end{pmatrix} \,.
\end{equation}
For a given Fermi energy \( \epsilon_F\), the spin splitting is given by \(\Delta=2\lambda k_F\), where \(k_F=\sqrt{2\epsilon_F}\) is the Fermi wavenumber (energy is measured in units of \( \hbar^2/ma^2 \)). In the chiral base the velocity operator \(v_x=\I\,[H,x]\) is given by,
\begin{equation}
v_{x}=\left(
\begin{array}{cc}
 (k-\lambda) \cos \varphi & -\I\lambda \sin \varphi  \\
 \I \lambda \sin \varphi  & (k+\lambda) \cos \varphi 
\end{array}
\right)\,,
\end{equation}
\(j_x=-v_x\) is the current in the \(x\) direction, and the spin current operator \(j^z_y=(1/4)\{\sigma_z,v_y\}\) is,
\begin{equation}
j^{z}_{y}=\left(
\begin{array}{cc}
 0 & -\frac{k}{2}  \sin \varphi  \\
 -\frac{k}{2} \sin \varphi  & 0
\end{array}
\right)\,.
\end{equation}
Note that the velocity operator, at variance with the spin current operator, explicitly depends on the spin-orbit coupling constant. We use the conventional definition of the spin current operator \(j^z_y\), but other choices are possible (in the presence of spin-orbit coupling it is not a conserved quantity).\cite{Shi-2006ys} 

In the presence of isotropic magnetic impurities, the scattering strength is characterized by a spin-flip relaxation time
\begin{equation}
\label{eqn.tau}
\frac{1}{\tau} = 2 \pi N_F c J_s^2\approx c J_s^2
\end{equation}
where \(c\) is the concentration of impurities (per unit area), and \(N_F\) is the density of states at the Fermi surface; it approaches the value \(N_F\approx 1/2\pi\) when the Fermi energy is much larger than any other energy in the system (\(N_F\approx m/2\pi\hbar^2\) in dimensional units).\cite{Moca-2007tg,Gorini-2008kx}

\subsection{Kubo conductivity}
\label{subsec.kubo}

\begin{figure}
\centering%
\includegraphics[width=0.48\textwidth]{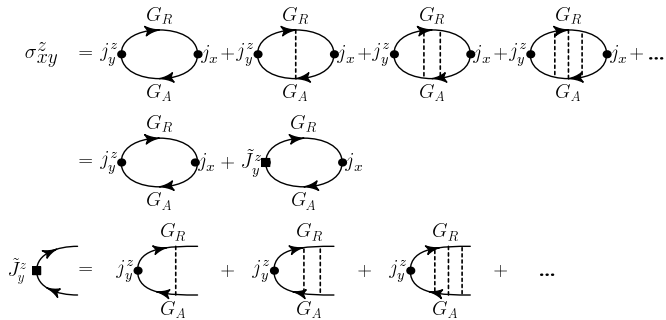}
\caption{(Color online) The conductivity is calculated in the non-crossing approximation, as a zero order loop with disorder averaged Green's functions, plus the ladder diagrams. The ladder diagrams are calculated by renormalization of the spin current vertex.}
\label{fig.loops}
\end{figure}

\begin{figure}
\centering%
\includegraphics[width=0.48\textwidth]{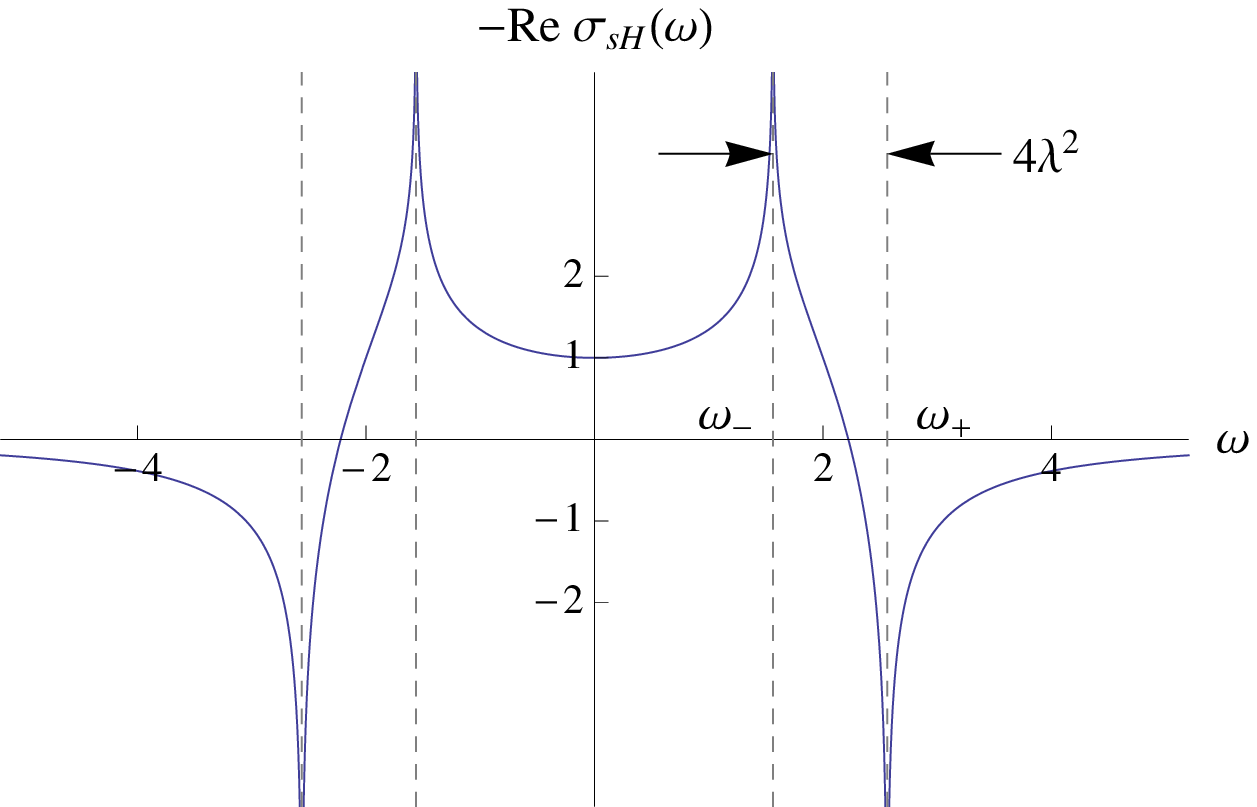}\\
    \includegraphics[width=0.48\textwidth]{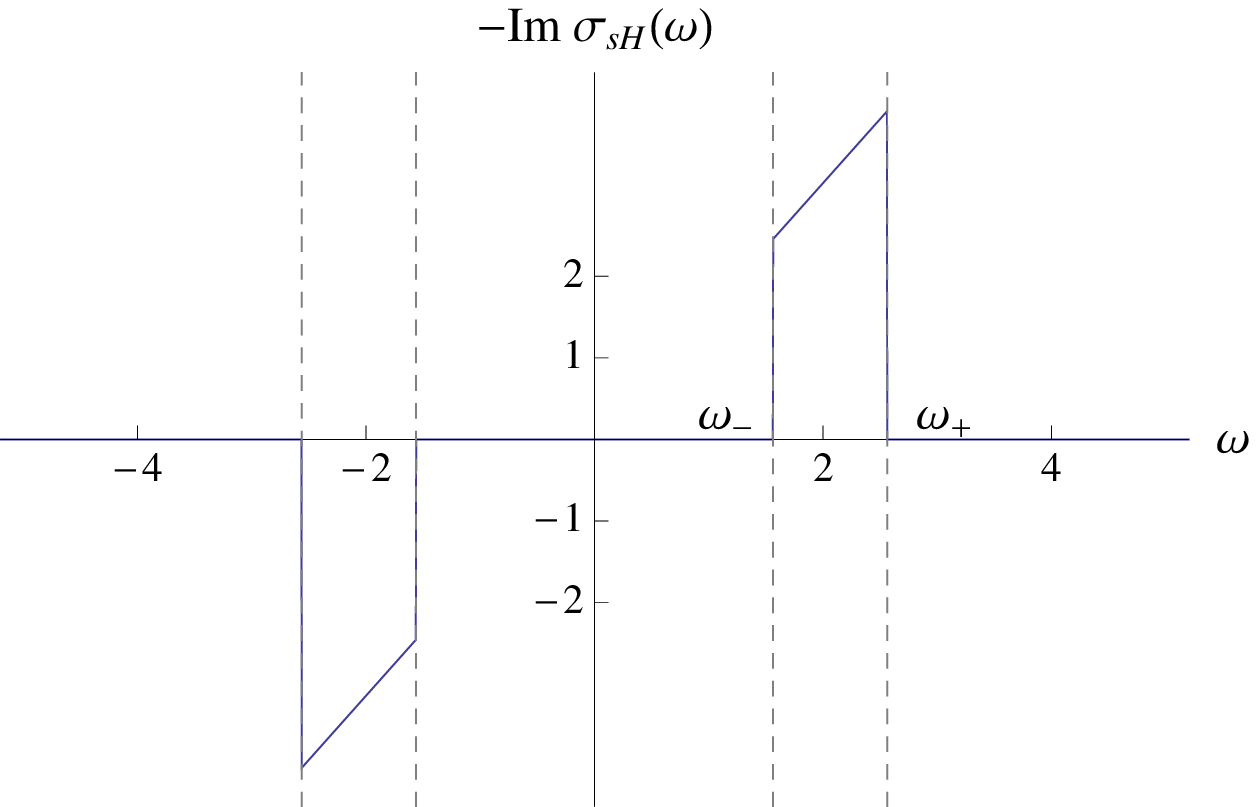}
\caption{(Color online) Real and imaginary parts of the intrinsic spin Hall conductivity for a pure system (\protect\ref{eqn.anaRe})-(\protect\ref{eqn.anaIm}). The four branches  \(\pm\omega_\pm\) are indicated by vertical dashed lines, their separation is \( |\omega_+ + \omega_-| = 4 \lambda^2 \) (arrows). }
\label{fig.exact}
\end{figure}

The spin Hall conductivity can be calculated in the framework of linear response theory,\cite{rammer1998quantum} using the following Kubo formula,\cite{Schliemann-2004fk}
\begin{equation}
\label{eqn.kubo}
\sigma_{sH}=\sigma^z_{xy}(\omega) = \frac{-\I}{L^2} \sum_{\alpha, \alpha '}
   \frac{f_{\alpha}- f_{\alpha '}}
   {\epsilon_{\alpha} - \epsilon_{\alpha '}} \,
   \frac{\langle \alpha | j^{z}_y | \alpha ' \rangle 
   \langle \alpha ' | j_x |\alpha \rangle}
   {\epsilon_{\alpha} - \epsilon_{\alpha '} + \omega + \I o} 
   \,,
\end{equation}
where \(\omega\) is the frequency of the applied external electric field, \(\I o\) is a small imaginary energy added to ensure causality, \(L^2\) is the system's area, \(\epsilon_{\alpha}\) and \(|\alpha\rangle\) are eigenvalues and eigenstates of \(H\), respectively, and \(f_\alpha=f(\epsilon_{\alpha})\) is the Fermi energy function. Formula (\ref{eqn.kubo}) is well suited for numerical computations of the tight-binding model\cite{Nomura-2005fk,Sheng-2005kl} (see Sec.~\ref{sec.numerics} below); for the conduction band model (\ref{Hcont}) it is convenient to express it in terms of retarded \(G^R\), and advanced \(G^A\), Green functions,\cite{Dimitrova-2005uq,Chalaev-2005fk}
\begin{align}
\label{eqn.sigma(G)}
\sigma^z_{xy} (\omega) = & \frac{1}{\omega} 
    \left\langle\mathrm{Tr} \left[ 
    j_y^{z} (\bm{k}) G^R (\epsilon+\omega,\bm{k}) 
    j_x (\bm{k}) G^< (\epsilon, \bm{k}) \right. \right.  
\nonumber \\
    & + \,\left. \left. 
    j_y^{z} (\bm{k}) G^< (\epsilon+\omega,\bm{k}) 
    j_x (\bm{k}) G^A(\epsilon, \bm{k}) 
    \right] \right\rangle,
\end{align}
where 
\[
\mathrm{Tr}[\dots]=\int \frac{d^2 \bm{k}}{(2 \pi)^2} 
    \int \frac{d\epsilon}{2 \pi} \mathrm{Tr}_s \,(\dots)
\]
\(\mathrm{Tr}_s\) is the trace over the spin index, \(\langle\dots\rangle\) denotes the average over the spatial disorder and magnetic moment orientations, and \( G^< (\epsilon) = f(\epsilon) (G^R(\epsilon)-G^A(\epsilon)) \). Calculation of ({\ref{eqn.sigma(G)}) can be done using the diagrams of Fig.~\ref{fig.loops}, where solid lines represent the disorder averaged Green functions, \(\langle G^{R,A}(\epsilon,\bm{k})\rangle\):\cite{Wang-2007uq}
\begin{equation}
\label{eqn.GR}
G^R_k (\epsilon)=\begin{pmatrix}
    \displaystyle\frac{1}{\epsilon-\epsilon_{-}(k) + \frac{\I }{2\tau}} & 0 \\
    0 & \displaystyle\frac{1}{\epsilon-\epsilon_{+}(k) + \frac{\I }{2\tau}}
    \end{pmatrix}
\end{equation}
and 
\begin{equation}
\label{eqn.GA}
G^A_k (\epsilon)=\begin{pmatrix}
    \displaystyle\frac{1}{\epsilon-\epsilon_{-}(k) - \frac{\I }{2\tau}} & 0 \\
    0 & \displaystyle\frac{1}{\epsilon-\epsilon_{+}(k) - \frac{\I }{2\tau}}
    \end{pmatrix}
\end{equation}
both written in the chiral basis. The perturbation expansion of Fig.~\ref{fig.loops} contains the lowest order one-loop contribution (the first diagram) and higher order one-loop contributions, containing disorder averaged terms, from which we only retain the non-crossing diagrams (ladder approximation). 

Before proceeding with the computation of these diagrams, we 
present the spin Hall conductivity in the clean limit, for which the Kubo formula can be calculated exactly. The real part of \(\sigma_{sH}\) is,
\begin{equation}
\label{eqn.anaRe}
\mathrm{Re}\,\sigma_{sH}(\omega)=\frac{-1}{8\pi}\left[1-\frac{\omega}{8\lambda^2}
   \log \left|\frac{
      \omega(\omega+4\lambda^2)-\Delta^2}{
      \omega(\omega-4\lambda^2)-\Delta^2}
      \right|\right],
\end{equation}
which has branch cuts at the roots \(\pm\omega_\pm\) with \(\omega_\pm=2\lambda^2 \pm \sqrt{4\lambda^4+\Delta^2}\) (Fig.~\ref{fig.exact}). This formula coincides with the result of Ref.~\onlinecite{Erlingsson-2005fk} in the small frequency limit. The imaginary part is given by
\begin{equation}
\label{eqn.anaIm}
\mathrm{Im}\,\sigma_{sH}(\omega)=\frac{-\omega}{64\lambda^2}
    \Theta\big[1-\frac{
    (\omega^2-\Delta^2)^2}
    {(4\lambda^2\omega)^2}\big]
\end{equation}
where \(\Theta\) is the Heaviside function (see Fig.~\ref{fig.exact}, bottom). 

For the evaluation of \(\sigma=\sigma^{z}_{xy}\), using the diagram expansion,
\[
\sigma_{sH}(\omega)=\sigma_{sH}^0(\omega)+\sigma_{sH}^L(\omega)\,,
\]
of Fig.~\ref{fig.loops}, we consider the weak scattering limit \(1/\tau,\,\Delta\ll \epsilon_F\), in which the products \(G^RG^R\) or \(G^AG^A\), of like Green functions are small compared to the product \(G^RG^A\):
\begin{equation}
\label{eqn.GrGa}
\sigma^z_{xy} (\omega) =  \big\langle \mathrm{Tr} 
   \big[\frac{f(\epsilon + \omega) - f(\epsilon)}{\omega}
   j_y^{z}
   G^R (\epsilon+\omega)j_x  
   G^A(\epsilon) \big] \big\rangle ,
\end{equation}
where we omitted the momentum variable \(\bm{k}\). Moreover, we consider the low frequency limit \(\omega\ll \epsilon_F\), so the difference of Fermi distributions is different from zero only in a small neighborhood of \(\epsilon_F\). The calculation of the first digram is straightforward,
\begin{align}
\label{eqn.cleansW}
\sigma_{sH}^{0}(\omega) &= \frac{-1}{8\pi} 
    \left[ 1 - \frac{(1- 
    \I \tau \omega)^2}{(1-\I \tau \omega)^2 + \Delta^2 \tau^2 } \right]
\nonumber\\
    &= \frac{-1}{8\pi}\frac{\Delta^2 \tau^2}{\Delta^2 \tau^2+w^2}\,,\quad
    w=1-\I \tau \omega
\end{align}
where the first term gives the intrinsic contribution (\(\tau\gg1\)) to the spin Hall conductivity (\(-e/8\pi\) in dimensional units). To this order, for which electron-impurity correlations are neglected, we obtain the same result as in the non-magnetic case.\cite{Chalaev-2005fk}

\begin{figure}
\begin{centering}
\includegraphics[width=\linewidth]{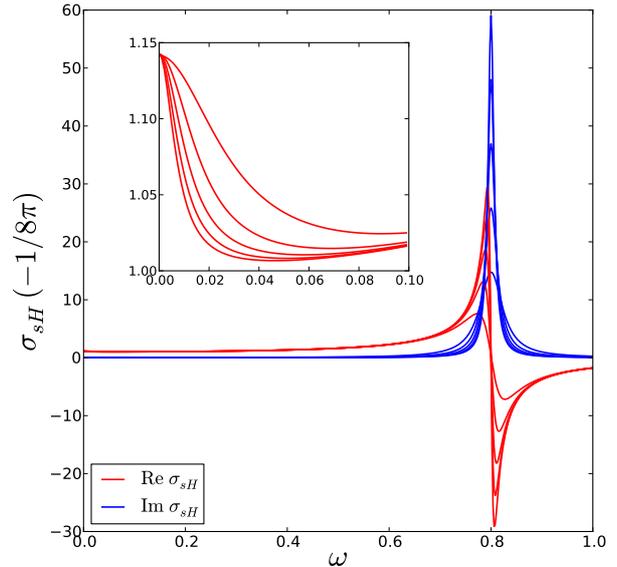}
\end{centering}
\caption{(Color online) Dynamical conductivity \(\sigma_{sH}\) as a function of \(\omega\), for \( \tau \) varying between \(40\) and \( 190\) (light to dark lines). (Inset) As \( \omega \rightarrow 0 \) the real part tends to \(1/8\pi\) for \(\omega>1/\tau\), but for \(\omega<1/\tau\) it increases up to \( 8/7 \approx 1.14\). }
\label{fig.s0sL}
\end{figure}

\begin{figure}
\centering%
\includegraphics[width=\linewidth]{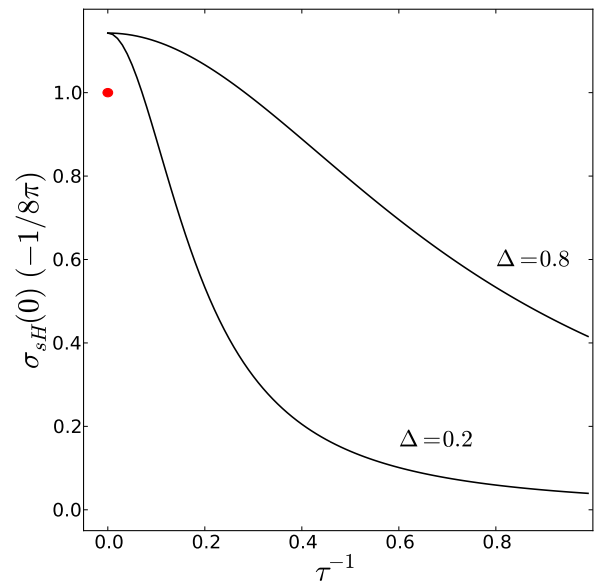}
\caption{(Color online) Static spin Hall conductivity as a function of the disorder strength from Eq.~(\protect{\ref{eqn.sigma}}), for two values of \(\Delta\). The dot at \(\tau^{-1}=0\) corresponds to the clean static limit.}
\label{fig.s0tau}
\end{figure}

Next, we compute the vertex correction (ladder diagrams, and third line in Fig.~\ref{fig.loops}) in the same approximation, where the Fermi energy \(\epsilon_F\), is much larger than any other energy scale:
\begin{equation}
\label{eqn.sLexpression}
\sigma_{sH}^{L}(\omega) =  \langle \mathrm{Tr} [ 
    J_y^{z}(\omega) G^R (\epsilon_F+\omega)  j_x G^A(\epsilon_F) 
    ] \rangle
\end{equation} 
where the trace now runs over momentum and spin, and \(J_y^{z} =J_y^{z} (\omega,\bm{k}) \) is the corrected spin current vertex, which is a function of the energy \(\omega\)). The algebraic equation that determines the corrected vertex, the third line of Fig.~\ref{fig.loops}, is conveniently expressed in the spin basis, 
\begin{align}
\label{eqn.selfconsistentvertex}
J_y^{z} = &\,c J_s^2  \int \frac{kdk d\varphi}{(2\pi)^2}
   \int\frac{\sin \theta d\theta d\phi}{4\pi}
   \begin{pmatrix}
       \cos \theta & \E^{- \I \phi}\sin \theta  \\
       \E^{\I \phi }\sin \theta & -\cos \theta 
   \end{pmatrix} 
\nonumber \\
    &  \times \, U(\varphi)G^A (0, \bm{k}) U^{\dagger}(\varphi) 
    \left[ j^{\sigma_z}_y + J_y^{z} \right] 
\\
    & \times  \, U(\varphi)G^R(\omega,\bm{k})U^{\dagger}(\varphi) 
    \begin{pmatrix}
        \cos \theta & \E^{- \I \phi} \sin \theta \\
        \E^{\I \phi }\sin \theta & -\cos \theta 
    \end{pmatrix}, 
\nonumber
\end{align}
where we used the unitary matrix (\ref{eqn.u}); we note that \(J_y^{z}\), in the spin basis, has the structure,
\begin{equation}
J_y^{z} = 
    \begin{pmatrix}
        0 & J_{\uparrow \downarrow} \\
        J_{\downarrow \uparrow} & 0 
    \end{pmatrix}\,, 
\end{equation}
with \(J_{\uparrow \downarrow}=J_{\downarrow \uparrow}\) given by,
\begin{equation}
\label{eqn.vertexJ}
J_{\uparrow \downarrow} = \frac{\I k_F \Delta \tau w}
    {2 w^2 + 2(7 w + \I \tau \omega)(\Delta^2 \tau^2 + w^2) }\,.
\end{equation}
Introducing this expression into (\ref{eqn.sLexpression}) and computing the trace, results in the ladder contribution,
\begin{equation}
\label{eqn.sLadder}
\sigma^{L}_{sH} (\omega)= \sigma^{0}_{sH}(\omega) 
    \frac{\Delta^2 \tau^2 + 4 \I \tau \omega }
    {w^2 + ( 7w+ \I \tau\omega)(\Delta^2 \tau^2 + w^2)}\,.
\end{equation}
Collecting the two contributions (\ref{eqn.cleansW}) and (\ref{eqn.sLadder}) we finally obtain the dynamical spin Hall conductivity in the presence of magnetic impurities, in the weak scattering strength approximation,
\begin{align}
\label{eqn.sigma}
\sigma_{sH} (\omega)= &\,
    \frac{-1}{8\pi} \frac{\Delta^2 \tau^2}{\Delta^2 \tau^2+w^2}
    \bigg[1 + \, 
\nonumber \\
    & +\, \frac{\Delta^2 \tau^2 + 4 \I \tau \omega }
    {2w^2(4-3\I \tau \omega) + \Delta^2 \tau^2( 7- 6\I \tau\omega)}
    \bigg]\,.
\end{align}
This result is similar to the one obtained in Ref.~\onlinecite{Gorini-2008kx}, although it has an extra term \(4 \I \tau \omega\) in the numerator of the last term; this difference, that does not change the limiting behavior for small frequency or large scattering times, can be attributed to the different approximation methods used. We show in Fig.~\ref{fig.s0sL} the real and imaginary parts of \(\sigma_{sH}\)
for different values of the disorder scattering time, as a function of the frequency. The real part of \(\sigma_{sH}\), an even function of the frequency, has a sharp change of sign at \(\omega=\Delta\) in the clean limit, that smoothes out with finite \(\tau\). The imaginary part shows, at the same frequency, a peak that broadens with increasing disorder. Comparing with Fig.~\ref{fig.exact}, we note that the in the weak scattering limit the splitting \(\omega_-+\omega_+\) is neglected.

The static spin current, as deduced from the real part of (\ref{eqn.sigma}) at \(\omega\rightarrow0\),
\begin{equation}
\sigma_{sH}(0)=\frac{-1}{8\pi}\frac{8\Delta^2 \tau^2}{8+7\Delta^2 \tau^2}\,,
\label{s0s}
\end{equation}
is shown in Fig.~\ref{fig.s0tau}. The continuous band model predicts an enhancement of the spin conductivity near the clean limit \(\tau\gg \Delta^{-1}\), and a slow reduction for smaller scattering times.
 
\subsection{The clean limit of \(\sigma_{sH}\)}
\label{subsec.limits}

\begin{figure}
\centering%
\includegraphics[width=\linewidth]{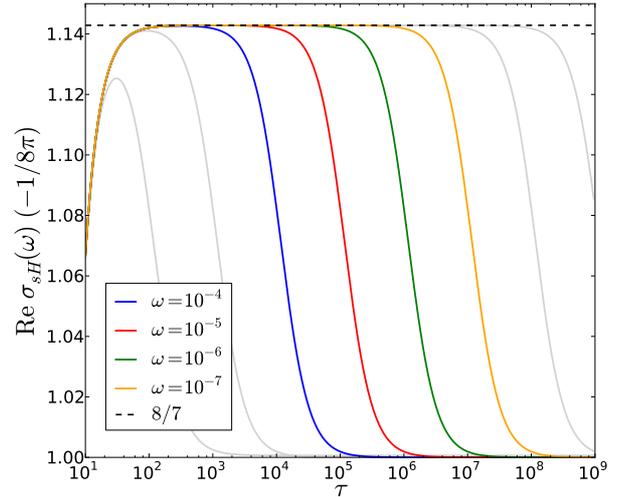}
\caption{(Color online) Convergence of the conductivity for values of  \( \omega \) near zero, ranging from \( 10^{-3} \) to \( 10^{-9} \). The expression of  \( 8 \pi Re \left[ \sigma (\omega)\right] \)  for \( \omega = 10^{-n} \) converges to zero only for  \( \tau > 10^{n+1}\). The horizontal line in the graph corresponds to the numerical factor \( 8/7 \).  }
\label{fig.limits}
\end{figure}

The imaginary part of \(\sigma_{sH}\) is an odd function of the frequency, and thus it must vanish at zero. Moreover, the real part, contrary to the non-magnetic case, does not vanish in the clean limit. For \(\tau \rightarrow \infty\) we find,
\begin{equation}
 \lim_{\tau \rightarrow \infty} \sigma_{xy}^z (\omega) =  \frac{-1}{8 \pi} \frac{\Delta^2}{\Delta^2 - \omega^2}\,;
 \label{eqn.cleansW0}
\end{equation}
if now we put \(\omega \rightarrow 0\) we recover the intrinsic value of the static spin Hall effect, as might be expected.\cite{Gorini-2008kx} However, the dynamical spin conductivity near the origin is singular,\cite{Arii-2007kx} as can be inferred from the behavior of \(\mathrm{Re}\,\sigma_{sH}\) near \( \omega=0\) (inset in Fig.~\ref{fig.s0sL}). Indeed, we observe that convergence to the static limit is not uniform: the larger the value of \(\tau\), the steeper the slope at the origin. This is related to the fact that the limits, \(\tau \rightarrow \infty\) and \( \omega \rightarrow 0\) do not commute:
\begin{equation}
 \lim_{\tau \rightarrow \infty} \sigma_{xy}^z (0) =  \frac{-1}{8 \pi} \frac{8}{7}\,,
 \label{eqn.cleansW-tw}
\end{equation}
as seen from formula (\ref{s0s}). Various different results for the clean limit of \(\sigma_{sH}(0)\) were reported.\cite{Inoue-2006fk,Wang-2007uq,Gorini-2008kx} The reason can be attributed to the fact that the static value \(\sigma_{sH}(0)\) cannot be computed at \(\omega=0\), for \(\tau\) tending to infinity. The order of limits starting with \(\omega \rightarrow 0\), would violate the causality condition contained in the linear response formula.

We present in Fig.~\ref{fig.limits} a sequence showing the spin conductivity as a function of the scattering time for the series \( \omega_n = 10^{-n} \), with \(n=2,\dots,9\). We observe that, for a fixed value of the frequency \(\omega_n\), convergence to \(-1/8\pi\) is reached for \(\tau>\tau_{n+1}\) (where \(\tau_n=10^n\)). Therefore, the limit \( \omega \rightarrow 0 \) does not converge for any finite value of \( \tau \). In fact, one can erroneously interpret the transitory value of \(8/7\) as being the static spin conductivity.\cite{Wang-2007uq} 

We conclude that in the presence of magnetic impurities, and in the weak disorder approximation, the static spin conductivity jumps from its intrinsic value to a value about \(8/7\) for finite disorder, and then it smoothly decreases when the disorder increases.

\section{The tight binding model and numerical results}
\label{sec.numerics}
We will now go beyond the effective mass model and the perturbation approximation in the disorder strength, to compute the spin conductivity. We introduce a discrete tight-binding model to investigate, in particular, the influence of a finite band width on the spin transport. The Hamiltonian is defined on a finite square lattice of size \(L \times L\) and lattice constant \(a\), with Rashba spin-orbit coupling \(\lambda\), and exchange interaction on impurities \(J_s\),
\begin{align}
\label{eqn.Hdiscrete}
H = - & t  \sum_{\langle i,j \rangle} c_i^{\dagger} c_j\, + 
    \frac{\lambda}{a} \sum_i (
    \I c_i^{\dagger} \sigma_x c_{i+\hat{y}} - 
    \I c_i^{\dagger} \sigma_y c_{i+\hat{x}} ) 
\nonumber \\ 
    &+ \,J_s \sum_{i \in I} c^{\dagger}_i \hat{\bm n}_i \cdot \bm {\sigma} c_i, 
\end{align}
where the hermitian conjugate terms of the first two sums are implicit;  \(c_i^\dagger = (c^\dagger_{i\uparrow}\; c^\dagger_{i\downarrow})\) is the creation operator at site \(i\), of spin up \(\uparrow\) or down \(\downarrow\), and \(c_i\) is the corresponding annihilation operator. We denoted \(\hat x\) and \(\hat y\) the two orthogonal directions in the lattice.  The kinetic term is limited to nearest neighbours and its bandwidth is \(4t\). This model is a straightforward generalization of (\ref{Hcont}), with discrete kinetic, spin-orbit and exchange terms. In the following we take \(t =1\) as the energy unit, and \(\hbar=a=e=1\) as before.

The computation of the Kubo formula (\ref{eqn.kubo}) needs in principle the knowledge of the whole spectrum and the full set of eigenvectors of \(H\). One can avoid this numerically expensive task, by reducing the formula to a trace, which can be evaluated with a stochastic method, in any basis. Specifically, we adopted a pseudospectral implementation of the kernel polynomial method.\cite{Weisse-2008uq} 

Because the real and imaginary parts of the spin conductivity (\ref{eqn.kubo}) are related by the Kramers-Kronig relations, it is enough to compute the optical spin conductivity \(\sigma_{o}=\mathrm{Im}\,\sigma_{sH}\), defined by,
\begin{equation}
\label{eqn.sigma(J)}
\sigma_{o}(\omega)=\frac{1}{L^2} \sum_{\alpha,\alpha'}J^z_{xy}(\alpha,\alpha')\frac{f_{\alpha} - f_{\alpha'}}{\omega}\delta(\omega-\epsilon_{\alpha} + \epsilon_{\alpha'})\,,
\end{equation}
where
\[
J^z_{xy} (\alpha, \alpha ')= \mathrm{Im} \left[ \langle \alpha | j^{z}_y | \alpha ' \rangle \langle \alpha ' | j_x |\alpha \rangle \right],
\] 
can be written in the form,
\begin{equation}
\label{eqn.jzxydelta}
\sigma_{o} (\omega)= \frac{1}{ \omega}  \int d\epsilon \ j(\epsilon + \omega, \epsilon) \left[ f(\epsilon) - f(\epsilon + \omega) \right]\,,
\end{equation}
using the definition,
\begin{equation}
\label{eqn.matrixj}
j(\epsilon, \epsilon ') = \frac{1}{L^2} \sum_{\alpha, \alpha '} J^z_{xy} (\alpha, \alpha ') \delta(\epsilon - \epsilon_{\alpha}) \delta(\epsilon '  - \epsilon_{\alpha '}).
\end{equation}
The advantage of this representation is that the matrix \(j(\epsilon,\epsilon')\) is a function of energies only which can be expanded in Chebyshev polynomials products \(T_n(\epsilon)T_m(\epsilon')\) with expansion moments \(\mu_{nm}\), which are easily computed using a stochastic trace evaluation method. We now briefly detail the algorithm. 

The first step is to scale and translate the energy, in such a way that the Hamiltonian spectrum is mapped into the interval \(\epsilon \in [-1,1]\). The series expansion of (\ref{eqn.matrixj}) can then be written,
\begin{align}
j(\epsilon,\epsilon') & = \sum_{n,m=0}^{\infty}  
   \frac{\mu_{nm} h_{nm}  T_n(\epsilon) T_m(\epsilon') }{
   \pi^2 \sqrt{1-\epsilon^2} \sqrt{1-\epsilon'^2} } \nonumber \\
   & \approx \sum_{n,m=0}^{M-1}  
   \frac{\mu_{nm} h_{nm} g_n g_m T_n(\epsilon) T_m(\epsilon') }{
   \pi^2 \sqrt{1-\epsilon^2} \sqrt{1-\epsilon'^2} }\,,
\label{eqn.chebyshevexp}
\end{align}
where \(M^2\) is the number of moments we use, \(T_m\) is the Chebyshev polynomial of order \(m\), 
\[
h_{nm} = \frac{2}{1+\delta_{n,0}} \frac{2}{1+ \delta_{m,0}}\,,
\]
and \(g_m\) is a filter that minimizes the Gibbs oscillations arising in truncating the series to a finite order (we use the Jackson kernel, see Ref.~\onlinecite{Weisse-2008uq}). 

The second step is to compute \(\mu_{nm}\). The presence of the radicals in the denominator allows to express the moments as usual scalar products,
\begin{equation}
\label{eqn.moments1}
\mu_{nm} = \int^{1}_{-1} \int^{1}_{-1} d\epsilon d\epsilon' 
    j(\epsilon,\epsilon') 
    T_n(\epsilon) T_m(\epsilon') 
\end{equation} 
which after rearranging,
\begin{eqnarray*}
\mu_{nm} &=& \frac{1}{L^2}  \mathrm{Im}  \sum_{\alpha,\alpha'}  
        \langle \alpha | j^{z}_y | \alpha ' \rangle 
        \langle \alpha ' | j_x |\alpha \rangle  
        T_n(\epsilon_{\alpha}) 
        T_m(\epsilon'_{\alpha'}) 
\\
    &=& \frac{1}{L^2} \mathrm{Im}  \sum_{\alpha,\alpha'}  
    \langle \alpha | T_n (H) j^{z}_y | \alpha'  \rangle 
    \langle \alpha ' | T_m(H) j_x |\alpha \rangle  \,,
\end{eqnarray*}
can be written as a trace formula,
\begin{equation}
\label{eqn.moments}
\mu_{nm} = \frac{1}{L^2} \mathrm{Im} \mathrm{Tr} \left[ 
        T_n(H) j^{z}_y T_m (H) j_x \right]
\end{equation}
In the third step, the trace is evaluated using an average over random states \(|r\rangle\),
\begin{equation*}
\label{eqn.moments1}
\mu_{nm} \approx \frac{1}{N_r} \sum_r \langle r|
        T_n(H) j^{z}_y T_m (H) j_x |r\rangle\,,
\end{equation*} 
where \(N_r\) is the number of random vectors needed for averaging. Next, one uses the recurrence relations between the Chebyshev polynomials to obtain the operator products to order \((n,m)\). In our case, even if the spin current operator is traceless, fast decrease of statistical errors and a rapid convergence of the algorithm is obtained. In order to optimize the Chebyshev recursion algorithm we adapted a pseudospectral procedure:
\begin{equation*}
H|r\rangle=\mathrm{IFFT}\big[H_0(\bm k)\mathrm{FFT}[|r\rangle]\big]+V|r\rangle\,,
\end{equation*}
where \(H_0\) contains the kinetic and spin-orbit terms, diagonal in the \(k\)-representation, \(V\) is the impurity potential, diagonal in the \(x\)-representation, and FFT, IFFT denote direct and inverse fast Fourier transforms, respectively. Once \(\mu_{m n}\) are known, we can efficiently obtain \(j(\epsilon,\epsilon')\) from formula (\ref{eqn.matrixj}) using \(M_\epsilon\) collocation points \(\epsilon_l = \cos[\pi(l+1/2)/M_\epsilon]\). One drawback of this method is that the \(\omega\)-dependent integral~(\ref{eqn.jzxydelta}), cannot be directly computed, since \(\epsilon_l + \omega\) is not necessarily a collocation point. One therefore has to perform an interpolation of some sort.

The last step is thus to compute the convolution integral (\ref{eqn.jzxydelta}).  We insert into this equation a \(\delta\)-function in its integral form, in order to separate the energy integrals. The expression for \(\sigma_{o}\) becomes,
\begin{equation*}
\sigma_{o} (\omega) =  \int \frac{dt}{2\pi} 
   \int d\epsilon d\epsilon' \frac{f_{\epsilon'} - f_\epsilon }{\omega}\,
   \E^{\I t (\epsilon - \epsilon' + \omega)} 
   j (\epsilon, \epsilon ')
\end{equation*}
We discretize the frequencies as well as the energies and the auxiliary time variable \(t\). The sampled \(\sigma_{o}\), finally takes the form of a Fourier transform
\begin{equation*}
\sigma_{o} (\omega_n)=\frac{1}{\omega_n} \sum_m 
    \E^{\I t_m \omega_n} F (t_m)\,,
\end{equation*}
where
\begin{equation*}
F (t_m)=\frac{1}{\omega_n} \sum_{i,j} 
   \E^{\I t_m (\epsilon_i - \epsilon_j)} 
   j(\epsilon_i, \epsilon_j)(f_i - f_j)\,,
\end{equation*}
and so \(\sigma_{o}(\omega_n)\) is straightforwardly evaluated by a fast Fourier transform of \(F\). From the expression of the optical spin conductivity we calculate its static value using the Kramers-Kronig relation,
\begin{equation}
\label{s0}
\sigma_{sH}(0)=\fint_{-\infty}^{\infty} \frac{d\omega}{\pi} \frac{\sigma_{o} (\omega)}{\omega}\,.
\end{equation}

The density of states,
\begin{equation}
\rho(\epsilon)=\sum_\alpha \delta(\epsilon-\epsilon_\alpha)\,,
\end{equation}
may be computed using the same method, and the expansion
\[
\rho(\epsilon) = \sum_{m=0}^{M-1} \frac{ \mu_m h_m g_m T_n(\epsilon)}{\pi \sqrt{1-\epsilon^2}} \,,
\]
where 
\[
\mu_m = \frac{1}{L} \mathrm{Tr}[T_n(H)]\, .
\]

In order to avoid finite size artifacts and to approach the thermodynamic limit faster, we use twisted boundary conditions: \(\Psi (x+L,y) = \mathrm{e}^{i \varphi_x} \Psi(x,y)\) and \(\Psi (x,y+L) = \mathrm{e}^{i \varphi_y} \Psi(x,y)\), for any state \(\Psi\). Physical quantities are averaged over many phase configurations \((\varphi_x , \varphi_y)\in [0,2\pi]^2\). This method has already been applied to determine the finite-size spin Hall conductivity, which in the case of non magnetic impurities, should vanish in the thermodynamic limit.\cite{Sheng-2005kl, Nomura-2005fk} In our case, the spin Hall conductivity being finite even at the thermodynamic limit, twisted boundary conditions constitute a useful tool to improve numerical convergence. For instance, the translational invariance of the disorder in the limit of infinite number of configurations, is recovered by averaging over twisted boundary conditions. 

\begin{figure}
\centering%
\includegraphics[width=\linewidth]{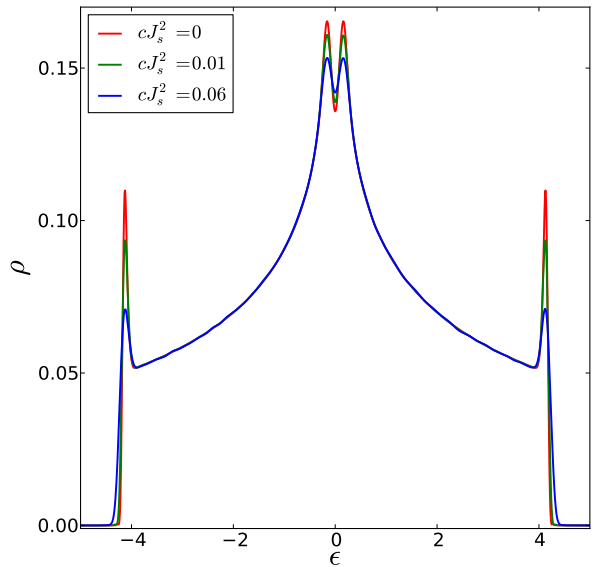}
\caption{(Color online) Density of states for \(c J_s^2= 0\) (clean system), and \(c J_s^2=0.01,\,0.06\); \(\lambda=0.4\). The spin-orbit interaction is responsible for the splitting of the chiral states (central peaks). }
\label{fig.rho}
\end{figure}

\begin{figure}
\centering%
\includegraphics[width=\linewidth]{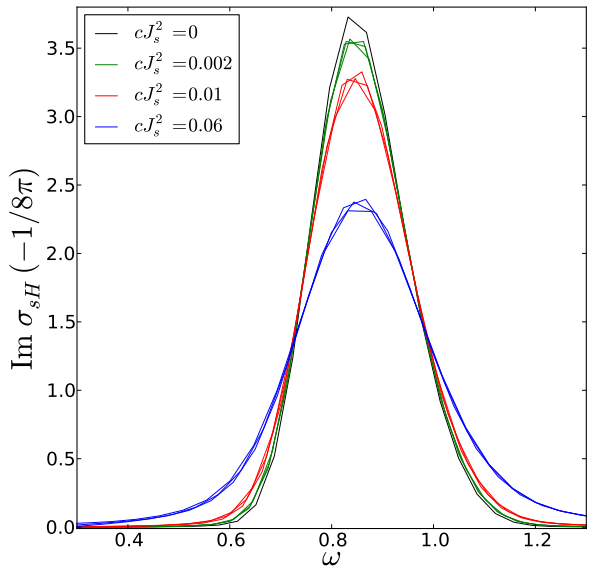}
\caption{(Color online) Optical spin conductivity as a function of the frequency, for different disorder strengths. For each \(\tau^{-1}=c J_s^2= 0.002,0.01,0.06\), three values of the concentration \(c=0.04,0.06,0.08\), were used; \(\epsilon_F=2\), \(\lambda=0.2\).}
\label{fig.sW_taudep}
\end{figure}

Let us make some remarks about the choice of the number of moments \(M\). Filtering the truncated Chebyshev expansion, introduces a broadening of sharp features that scales as \(1/M\), causing a reduction in the energy resolution. The optimal number of moments can be calibrated from the clean (\(J_s=0\)) case, and it turns out to be a few times the linear system size \( M = 2L,\, 4L\), therefore \(4L^2,\, 16L^2\) coefficients are involved in the \(j(\epsilon,\epsilon')\) expansion. We obtained a a well detailed density of states, without finite-size artifacts. This choice of numerical parameters, has been verified by comparing small samples averaged over twisted boundary conditions, with equivalent samples of greater size.

\subsection{Numerical results}
\label{subsec.sW}

In the following, we describe the properties of spin transport through the numerical computation of the density of states, and the static and optical spin conductivities. Simulations were performed on lattices of size \(64^2,\,128^2\), but higher sizes were used to test convergence. Typical numerical parameters are: \(\lambda = 0.2, \, 0.4\), in units of \(at\), \(c\) from \(2\%\) to \(10\%\), and \(\tau\) from \(5\) to \(500\), in units of \(\hbar/t\). 

We will limit the range of parameters in order to preserve the qualitative shape of the clean density of states, avoiding for instance an impurity band regime (strong disorder regime). This choice should allow us to compare the tight-binding model to the analytical results of the preceding section, and to determine its validity. One observes in Fig.~\ref{fig.rho} that, within the range \(\tau^{-1}=0,\dots,0.06\), the density of states \(\rho(\epsilon)\) remains close to its clean limit, and the spin-orbit splitting, proportional to \(\lambda\), fades out with stronger scattering (to enhance the effect of the spin-orbit coupling, we put \(\lambda=0.4\)).

\begin{figure}
\centering%
\includegraphics[width=\linewidth]{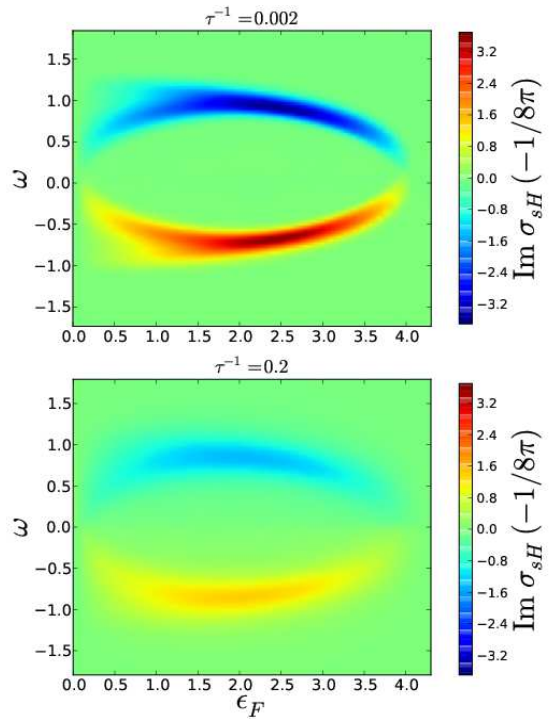}
\caption{(Color online) Optical spin conductivity as a function of frequency and \(\epsilon_F\), for two different disorder strengths \(\tau^{-1}=0.002,\,0.2\). As the disorder increases, the optical conductivity decreases and spreads out over \(\omega\). }
\label{fig.sW_Efwdep}
\end{figure}

The spin conductivity depends on the Fermi energy \(\epsilon_F\), the spin-orbit coupling \(\lambda\), and  in principle, the concentration \(c\) and the exchange constant \(J_s\). In fact, the scattering off impurities is characterized by the sole parameter \(\tau^{-1}=cJ_s^2\), at least in the weak disorder limit \(\tau^{-1}\ll 1\), hence we would expect that \(\sigma_o=\sigma_o(\omega;\tau)\). In Fig.~\ref{fig.sW_taudep} we represented the optical spin conductivity as a function of the frequency, for different sets of pairs \(c\) and \(J_s\), giving the same values of \(\tau\). We chose the Fermi energy at the middle of the band, \(\epsilon_F=2\). We observe that the curves coincide according to their \(\tau\) values, confirming that in the regime studied, the relevant parameter is the product \(cJ_s^2\). The maximum of \(\sigma_o\) corresponds to \(\omega\approx \Delta\) as in the continuous case, and is almost independent of \(\tau\) (compare with Fig.~\ref{fig.exact}). In Fig.~\ref{fig.sW_Efwdep}, we present \(\sigma_o\) as a function of the frequency and the Fermi energy. Its maximum appears at Fermi energies in the middle of the band, and follows a law in \(\omega\sim\sqrt{\epsilon_F}\), compatible with the estimation \(\omega\approx \Delta\) given above. The intensity of the peak decreased by a factor 2, for a factor of 100 in \(\tau\) (it is 500 for the top figure and 5 for the bottom one). The optical spin conductivity vanishes for small \(\omega\ll\Delta\), and large frequencies \(\omega\gg\Delta\). Therefore, the effect of increasing the disorder is, as expected, to broaden and damp the peak on the spectrum. However, at variance with the conduction band model for which the peak width is proportional to \(\tau^{-1} \), in the tight-binding model, the effect of a finite band width quantitatively change the scaling with the scattering time (the large Fermi energy limit is non longer applicable). 

The computation of the static spin conductivity (\ref{s0}), appears to be particularly sensitive to statistical errors, as already noted.\cite{Sheng-2005kl} A large set of boundary conditions and random impurities distributions, is necessary. Results are shown in Fig.~\ref{fig.s0_v0dep}, where \(\sigma_{sH}(0)\) is represented as a function of the Fermi energy and the disorder strength. The spin Hall conductivity is an odd function, it vanishes at \(\epsilon_F=0\) and outside the energy band, but is nearly constant over a large range of Fermi energies. Its slow fall off with disorder underlines the weakness of vertex corrections in the magnetic case, as shown by the analytical calculation of the previous section.\cite{Moca-2007tg} More importantly, we observe large fluctuations of the spin conductivity at low frequencies, depending on the sample distribution. Strong spin conductance fluctuations are known to arise at mesoscopic scales,\cite{Ren-2006fk} and, provided that scattering ensures ergodicity, their universal statistics can be related to that of random matrices.\cite{Bardarson-2007fk} We found a noticeable amplification of the fluctuations in the region \(\omega\lesssim\tau^{-1}\) that explains the large dispersion of \(\sigma_{sH}(0)\) values around its mean. In the inset of Fig.~\ref{fig.sW_taudep}, the integrand \(\sigma_o/\omega\) in (\ref{s0}), is represented together with error bars proportional to the sample to sample dispersion,
\begin{equation}
\label{deltas}
\Delta\sigma_{sH}(\omega)=\langle \sigma_{sH}(\omega)^2-
	\langle \sigma_{sH}(\omega)\rangle^2\rangle^{1/2}\,,
\end{equation}
where the angle brackets stand for the random potential averaging. The averaged static spin Hall conductivity, is then not enough to characterize the spin transport. In fact, even in the present weak disorder limit, where localization of the carriers can be neglected, we find a Gaussian distribution of the spin conductivity fluctuations, as illustrated in Fig.~\ref{fig.hist}. The standard deviation of the static spin conductivity \(\Delta \sigma_sH(0)/\langle \sigma_{sH}(0)\rangle \approx 0.1\), giving an incertitude of 10\% about the mean value.

\begin{figure}
\centering%
\includegraphics[width=\linewidth]{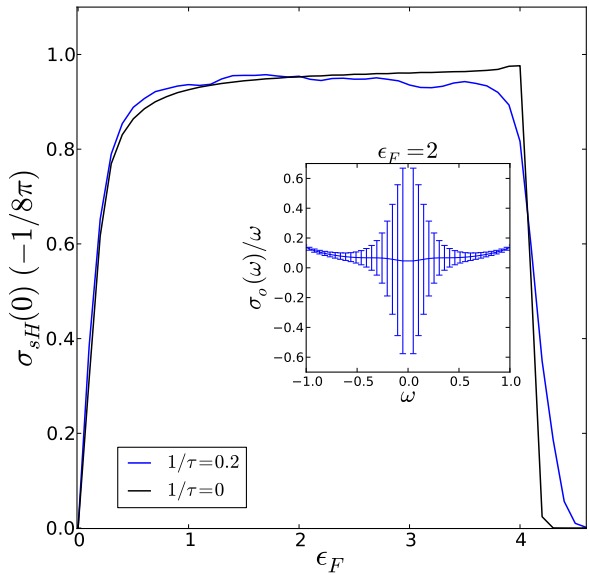}
\caption{(Color online) Static spin Hall conductivity as a function of \(\epsilon_F\), for disorder strengths \(\tau^{-1}=0,\,0.2\). (Inset) Integrand of (\protect\ref{s0}), used to compute \(\sigma_{sH}(0)\); the error bars represent the amplitude of \(\sigma_o(\omega)/\omega\) fluctuations near the zero frequency.}
\label{fig.s0_v0dep}
\end{figure}

\begin{figure}
\centering%
\includegraphics[width=\linewidth]{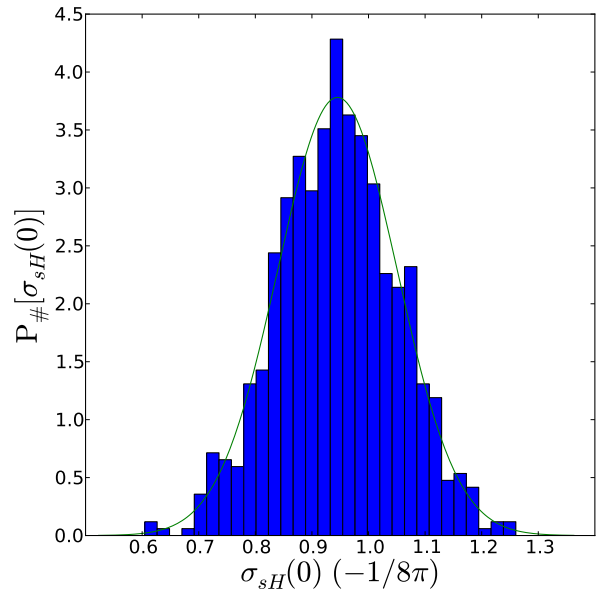}
\caption{(Color online) Normalized histogram of the static spin conductivity for \(\tau^{-1}=0.2\), showing a Gaussian distribution with mean value \(\langle \sigma_{sH}(\omega)\rangle=0.98\, [-1/8\pi]\) and  standard deviation \(\Delta \sigma_{sH}(0) =0.1\) (solid line).}
\label{fig.hist}
\end{figure}

\section{Conclusion}
\label{sec.conclusion}
In this paper we investigated the influence of magnetic impurities in the intrinsic spin Hall effect. A first model, amenable to analytical calculations, was introduced. We considered a simple one conduction band Hamiltonian including the effects of Rashba spin-orbit coupling and exchange interaction with randomly distributed magnetic moments. Using standard linear response methods we obtained, in the weak scattering and low frequency approximation, the expression of the dynamical spin conductivity as a function of the system's parameters (spin-orbit coupling, scattering time, Fermi energy). We emphasized the relevance of the order of the large scattering time and small frequency limits, and found that when frequency is going to zero, one must recover the spin conductivity of an ideal two dimensional electron gaz, \(\sigma=-e/8\pi\). However, an arbitrarily small amount of magnetic impurities will increase the static spin conductivity by a factor of about \(8/7\), showing the singular nature of the clean limit.

We also investigated, using numerical methods, the tight-binding generalization of the conduction band Hamiltonian. We adapted the kernel polynomial method using a pseudospectral evaluation of the Chebyshev recursion to compute the density of states and the optical spin conductivity. As a function of frequency, the spin Hall conductivity shows a peak at \(\omega\sim \lambda\sqrt{\epsilon_F}\),  that broadens and weakens with the increase of the disorder strength. These changes are determined by the combination \(\tau^{-1}\sim cJ_s^2\), that characterize the disorder strength.

Within the regime of weak disorder, that is in the absence of separate impurity bands, the deviation with respect to the clean system is small. The main effect of magnetic impurities in contrast with non-magnetic ones, is to preserve and reinforce the intrinsic mechanism that would otherwise be destroyed. In the finite band model, we observe an amplification of spin conductivity fluctuations at low frequencies, which are at the origin of a significant incertitude of the static \(\sigma_{sH}\) mean value. A detailed theory of the spin Hall conductivity in the presence of magnetic impurities should consider these sample to sample fluctuations, that could become the main physical effect in the description of spin transport in the strong disorder regime.

In spite of the absence of magnetic order (we treated the paramagnetic state), the sole presence of spin-flip processes, guarantees the emergence of a spin current proportional to the applied electric field. For weak disorder, and strong spin-orbit coupling, the intrinsic spin Hall effect is reinforced by multiple scattering interactions, described by the vertex correction to the spin current. Manipulation of the carrier's spin in a semiconductor impurities, is then possible by doping it with magnetic impurities and applying an electric field. This result naturally raises the question of the interplay between intrinsic spin Hall currents and ferromagnetic impurities as present in a diluted magnetic semiconductor.

\acknowledgments
We would like to thank A. Wei\ss e for his kind advice, and TM, AC for useful discussions.


%

\end{document}